\pdfoutput=1

\documentclass[fleqn,11pt]{article}

\usepackage{amsfonts,amsmath,amssymb}
\usepackage{latexsym}

\usepackage{amsfonts,amssymb,cite}
\usepackage{graphicx}

\def\noi{\noindent}
\def\jnumber#1#2{\thispagestyle{empty} \noi\unitlength=1mm
    \begin{picture}(178,10)
            \put(177,15){\llap{\large\it Grav. Cosmol. No.\,#1, #2}}
                    \end{picture}}

\newcommand{\Title}[1]{\noi {{\Large\bf #1}}\\[1ex]}

\def\Aunames#1{\noi{\bf #1}}
\def\au#1{${}^{#1}$}
\def\Addresses#1{\medskip\noi \protect
  \begin{description}\itemsep -3pt {\it #1} \end{description}}
\def\adr#1#2{\item[${}^{#1}$]{\it #2}}

\newcommand{\Abstract}[1]{\vskip 2mm \begin{center}
        \parbox{16.4cm}{\small\noi #1} \end{center}\medskip}

\def\email#1#2{\footnotetext[#1]{e-mail: #2}\addtocounter{footnote}{1}}

\def\nqq{\hspace*{-2em}}

\def\cm{\hspace*{1cm}}

\def\degC{\mbox{${}^\circ$C}}                
\usepackage{color}

\def\Acknow#1{\subsection*{Acknowledgments} #1}

\def\ConflictThey{\subsection*{Conflict of interest} 
  The authors declare that they have no conflicts of interest.}

\def\Jl#1#2{#1 {\bf #2},\ }

\def\ApJ#1 {\Jl{Astroph. J.}{#1}}
\def\CQG#1 {\Jl{Class. Quantum Grav.}{#1}}
\def\DAN#1 {\Jl{Dokl. AN SSSR}{#1}}
\def\GC#1 {\Jl{Grav. Cosmol.}{#1}}
\def\GRG#1 {\Jl{Gen. Rel. Grav.}{#1}}
\def\IJMPD#1 {\Jl{Int. J. Mod. Phys. D}{#1}}
\def\JETF#1 {\Jl{Zh. Eksp. Teor. Fiz.}{#1}}
\def\JETP#1 {\Jl{Sov. Phys. JETP}{#1}}
\def\JHEP#1 {\Jl{JHEP}{#1}}
\def\JMP#1 {\Jl{J. Math. Phys.}{#1}}
\def\NPB#1 {\Jl{Nucl. Phys. B}{#1}}
\def\NP#1 {\Jl{Nucl. Phys.}{#1}}
\def\PLA#1 {\Jl{Phys. Lett. A}{#1}}
\def\PLB#1 {\Jl{Phys. Lett. B}{#1}}
\def\PRD#1 {\Jl{Phys. Rev. D}{#1}}
\def\PRL#1 {\Jl{Phys. Rev. Lett.}{#1}}

\def\lal{&&\nqq {}}

\def\beq{\begin{equation}}
\def\eeq{\end{equation}}
\def\bear{\begin{eqnarray}}
\def\bearr{\begin{eqnarray} \lal}
\def\ear{\end{eqnarray}}
\def\earn{\nonumber \end{eqnarray}}

\def\yy{\\[5pt] {}}

\begin{document}
\twocolumn[
\jnumber{issue}{year}

\Title{Krizek--Somer Anthropic Principle and the Problem of\yy 
       Local Hubble Expansion}

\Aunames{Yurii V. Dumin\au{a,1} and Eugen S. Savinykh\au{2}}

\Addresses{
\adr a {Sternberg Astronomical Institute (GAISh),
        Lomonosov Moscow State University,\\
        Universitetskii prosp.\ 13, Moscow, 119234 Russia}
}


\Abstract
{According to the Krizek--Somer hypothesis~[New Astron. {\bf 17}, 1 (2012);
Grav.\ Cosmol. {\bf 21}, 59 (2015)], a biological evolution of the Earth
is possible only at certain values of the Hubble parameter,
because the increasing luminosity of the Sun should be compensated by the
increasing orbital radius of the Earth due to the local Hubble expansion,
thereby keeping the Earth's surface temperature sufficiently stable.
Here, we examine this hypothesis in light of the recent data on the surface
temperature on the early Earth, thereby imposing a few constraints on
the admissible values of the local Hubble parameter.
As follows from our analysis, the Krizek--Somer mechanism might be a valuable
tool to resolve the important geophysical and paleontological puzzles, but
the particular value of the local Hubble parameter is substantially affected
by the current uncertainties in our knowledge about the temperature and
other properties of the early Earth.}
\medskip

%

] 
\email 1 {dumin@pks.mpg.de, dumin@yahoo.com\\ \cm (Corresponding author)}
\email 2 {z-0000001@hotmail.com\\ \cm (Retired)}

\section{Introduction}

In general, the anthropic principle is formulated as a statement that
the emergence and development of life on the Earth is possible only
within a quite narrow range of the fundamental physical constants,
which---by a lucky coincidence---was realized in nature.
Such fundamental constants are usually claimed to be the masses of
elementary particles (electron and proton), the fine-structure constant,
\textit{etc.}
Sometimes, the dimensionality of our space is also included here.

An important and unexpected addition to this list was done a few years ago
by M.~Krizek and L.~Somer~\cite{Krizek_12,Krizek_15}, who showed that yet
another crucial factor for the existence of life on the Earth should be
the value of Hubble parameter, \textit{i.e.}, a relative expansion rate of
the Universe.
They drew this conclusion starting from the so-called faint young Sun
paradox, which was discussed in geophysics and astronomy since
1950s~\cite{Schwarzschild_58}.
Namely, as follows from the contemporary models of stellar evolution,
a luminosity of the Sun 4--4.5~billion years ago (when the Earth formed)
was about~25--30\,\% below its present-day value.
If one assumes that physical properties of the Earth's atmosphere remained
more or less permanent, then it can be easily found that all water on
the Earth would be frozen 2~billion years ago and earlier.
On the other hand, this fact evidently contradicts some geological data on
the existence of considerable volumes of liquid water on the Earth in that
period of time.
Moreover, it is even more important that freezing all water would made
impossible both the emergence of life on the Earth as well as its subsequent
development.

This problem was clearly recognized by the beginning of 1970s, first of all,
due to the works by C.~Sagan and G.~Mullen~\cite{Sagan_72,Sagan_77}.
As a plausible remedy, they suggested the so-called greenhouse effect,
\textit{i.e.}, keeping the outgoing infrared radiation from the planetary
surface by certain gases in its atmosphere.
Subsequently, a few other mechanisms to support the required thermal balance
of the Earth were proposed, such as a variable albedo of its surface,
the geothermal sources of energy, formation of the oil film on the surface of
ancient oceans, \textit{etc.}~\cite{Feulner_12,Khramova_24}.
Unfortunately, all these hypotheses suffer from the considerable uncertainty
in our knowledge of physical and chemical conditions on the early Earth.

A radically new approach to the problem of stabilization of temperature on
the Earth was put forward in the above-mentioned papers by Krizek and
Somer~\cite{Krizek_12,Krizek_15}, who suggested that a reduced solar
luminosity in the past could be compensated by the smaller planetary radii,
which subsequently expanded due to the local Hubble effect.
Particularly,
it was empirically found in paper~\cite{Krizek_15}
that at the local Hubble parameter~$ H_0^{\rm (loc)} = 0.52 \, H_0 $
(where $ H_0 \approx 67 $\,km/s/Mpc~was used as the standard intergalactic
value) the flux of solar energy received by the Earth turns out to be stable
with accuracy about~1\,\% throughout the entire period of 3.5~billion years
in the past and at a comparable time interval in the future.
This led them to the concept of ``ecosphere'' (\textit{i.e.}, a spherical
envelope about the Sun that is favorable for life), expanding in space
approximately with the Hubble velocity.
Unfortunately, the value of the local Hubble parameter two times smaller
than at the intergalactic scale looks rather suspicious.

Besides, despite an attractiveness of the above-stated idea, its key
requirement of the constancy of the solar radiation flux to the Earth's
surface (and, thereby, stability of the Earth's temperature) is by no means
self-evident.
Really, as follows from the recent paleochemical and paleobiological
studies and summarized, \textit{e.g.}, in paper~\cite{Gaucher_08},
the Earth's surface temperature in the period when the life emerged
might be considerably higher than its present-day value,
up to~$ 70{-}80\degC$.
As a result, it should be necessary to assume a higher value
of~$ H_0^{\rm (loc)} $, because the Earth would be closer to
the Sun in that time.

So, it is the aim of the present paper to refine the Kiizek--Somer mechanism
by taking into account the available data on temperature of the early Earth.
Thereby, it will be possible to impose a few constraints on the ``anthropic''
value of the Hubble parameter.

\section{Refinement of the Krizek--Somer Mechanism}

A thermal balance of the Earth can be described in the first approximation
by the equation:
\beq
\label{therm_bal}
\pi R^2 (1{-}A) K = 4 \pi R^2 (1{-}{\alpha}) \, \sigma \, T_{\rm s}^4 \, ,
\eeq
where
$ R $~is the Earth's radius,
$ T_{\rm s} $~is the temperature of its surface,
$ A $~is the Earth's albedo in the visible spectral range (which comprises
the most part of the arriving solar energy),
$ \alpha $~is the albedo of the Earth's atmosphere in the infrared range
(where thermal radiation leaves the Eath's surface),
$ K $~is the so-called solar constant, \textit{i.e.}, the flux of solar
radiation energy per unitary area of the Earth's surface, and
$ \sigma $~is the Stefan--Boltzmann constant.
The left-hand side of this equation
represents the flux of solar energy absorbed by the Earth's
cross-section~$\pi R^2 $,
while the right-hand side
describes the thermal infrared flux emitted from the entire surface
of the globe~$ 4 \pi R^2 $; both terms being corrected for the opacity
in the respective spectral ranges.

The above-mentioned ``solar constant'' should be actually the function of
time:
\beq
\label{solar_const}
K(t) = \frac{\displaystyle L(t)}{\displaystyle 4 \pi r^2(t)} \, ,
\eeq
because both the solar luminosity~$ L $ and the distance from Sun to the
Earth~$ r $ are time-dependent in the framework of Krizek--Somer model.

Next, let us assume that Hubble expansion at the sufficiently small scales
is produced only by the perfectly-uniform Dark energy or $ \Lambda $-term,
while the gravitational action by compact objects is described by the
ordinary Newtonian forces;
for more details about this approximation, see our earlier
papers~\cite{Dumin_08,Dumin_18}.
The second important simplification, immediately following from
the equation of state of the Dark energy, is that the temporal evolution of
the cosmological scale factor will be purely exponential, \textit{i.e.},
corresponding to the de~Sitter model.
Then,
distance from the Sun to the Earth
(which is actually the physical separation between the two points with
fixed co-moving Robertson--Walker coordinates~\cite{Misner_73})
should increase as
\beq
\label{expansion_rate}
r(t) = r_0 \exp \! \big( \! H_0^{\rm (loc)} t \big) \, ,
\eeq
where
$ r_0 $~is the Earth's orbital radius at the present time, and
$ H_0^{\rm (loc)} \! = \mbox{const} $.
(Additional discussion of this approximation will be given in
the concluding section of this paper.)

It can be easily found from formulas~(\ref{therm_bal})--(\ref{expansion_rate})
how the Earth's surface temperature~$ T_{\rm s} $ evolved in the past
(\textit{i.e.}, at negative~$ t $) at the specified rate of the local Hubble
expansion:
\beq
\label{T_t}
T_{\rm s}(t) = T_{{\rm s}0} \exp \! \big( \!{-} H_0^{\rm (loc)} t/2 \big)
  \bigg[\! \frac{\displaystyle L(t)}{\displaystyle L_0} \!\bigg]^{\! 1/4} ,
\eeq
where $ T_{{\rm s}0} \approx 15\degC $ and $ L_0 $~are the Earth's
surface temperature and solar luminosity at the present time (\textit{i.e.},
at $ t = 0 $).

The long-term evolution of the solar luminosity~$ L(t) $, in general, is
obtained by a cumbersome numerical integration of the equations of stellar
evolution.
However, the final result can be presented by a quite simple
formula~\cite{Gough_81}:
\beq
\label{L_t}
L(t) = L_0 \bigg( \! 1 - \frac{2}{5} \, \frac{t}{t_{\rm S}}
  \bigg)^{\!\!\! -1} ,
\eeq
where $ t_{\rm S} \approx 4.7{\cdot}10^9 $\,yr~is the age of the Sun.
Finally, inserting formula~(\ref{L_t}) into~(\ref{T_t}), we get the required
result:
\beq
\label{T_t-final}
T_{\rm s}(t) = T_{{\rm s}0} \exp \! \big( \!{-} H_0^{\rm (loc)} t/2 \big)
  \bigg[ 1 - \, \frac{2}{5} \, \frac{t}{t_{\rm S}} \bigg]^{\! -1/4} .
\eeq 

\begin{figure}
\centering
\includegraphics[width=8.2cm]{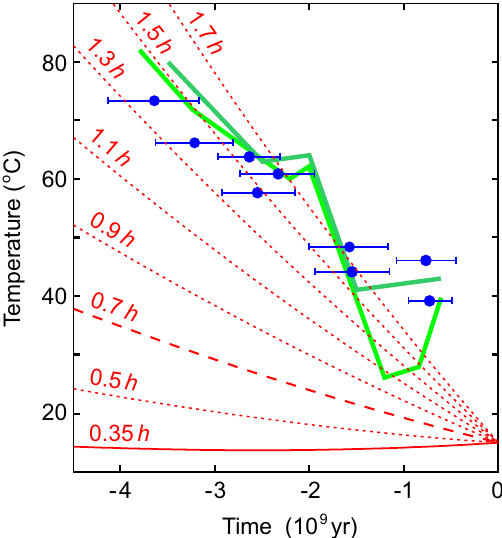}
\caption{\small 
The average surface temperature on the Earth~$ T_{\rm s} $ as function of
time~$ t $ in the past:
paleochemical (two thick green curves) and paleobiological data
(blue circles with error pars) \textit{vs.} the theoretical predictions
(thin red curves) at various values of the local Hubble parameter.}
\label{fig:T-t}
\end{figure}  

A series of temporal tracks~$ T_{\rm s}(t) $ at various values of the local
Hubble parameter are confronted with observational data in Fig.~\ref{fig:T-t}.
(For convenience, the values of~$ H_0^{\rm (loc)} $ are normalized to
$ h = 100 $~km/s/Mpc~$ \approx 3.2{\cdot}10^{-18} $\,s$ {}^{-1} $.)
Two thick green curves are the paleochemical data derived from the deposition
of various isotopes in the ocean, and the blue circles with error bars are
the paleobiological data based on modelling the ancient
proteins~\cite{Gaucher_08}.
The theoretical dependences~(\ref{T_t-final}) are drawn by the thin red
curves:
the dashed one corresponds to the standard intergalactic value~$ 0.7 \, h $;
the solid curve ($ 0.35 \, h $) refers to the original Krizek--Somer
proposition of permanent temperature on the Earth; and
the dotted curves correspond to a set of other values of the local Hubble
parameter.
It is seen that the observational data for the period earlier than 2~billion
years ago can be reasonably fit to the theoretical curves if
$ H_0^{\rm (loc)}{\approx}\,1.5\,h  $.

Therefore, the main conclusion following from this analysis is that---while
the original Krizek--Somer hypothesis required the local Hubble parameter
to be about two times smaller than the global one---the attempt of fitting to
the resent observational data on the Earth's surface
temperature~\cite{Gaucher_08} also requires a substantial deviation from
the global (intergalactic) Hubble parameter but in the opposite direction.
Namely, its local value should be about two times greater than the global one.
Besides, one can see significant outliers in the figure at the time interval
about one million years ago.
They might be caused either by a considerable greenhouse effect in the period
of active biological evolution (and the resulting change in the chemical
composition of the atmosphere) or just by the experimental errors, since
the scatter of data---both paleochemical and paleobiological---is rather
large.

\section{Discussion and Conclusions}

Referring to the viability of Krizek--Somer hypothesis and the respective
interpretation of the Hubble parameter as one of the ``anthropic'' factors,
it should be emphasized first of all that the possibility of local
(interplanetary) Hubble expansion \textit{per se} remains a very questionable
issue till now. Its study was initiated by G.C.~McVittie over 90~years
ago~\cite{McVittie_33}, and a few dozen of papers on that topic were published
in the subsequent decades; \textit{e.g.}, reviews~\cite{Bonnor_00,Dumin_16}.
The most of astronomers intuitively believed that a criterion for the onset
of Hubble expansion would be a gravitational binding of the system.
In other words, when the massive bodies become sufficiently heavy and begin
to move around each other in the confined orbits, they will no longer
experience the cosmological expansion.

However, this point of view was refutated by A.~Einstein and
E.G.~Straus~\cite{Einstein_45}, who have shown that Hubble expansion is
suppressed in the regions of low rather than high matter density, and it is
restored again when the density reaches its cosmological background value.
In other words, the cosmological expansion is a specific ``degree of freedom''
of the gravitational field, which is unrelated to the ordinary Newtonian
forces.
Really, the Hubble expansion exists even in the perfectly uniform Universe,
where there are no any gravitational forces at all.

The possibility of the local (\textit{e.g.}, interplanetary) Hubble expansion
looks especially plausible in the framework of Dark-energy-dominated models,
which became the dominant paradigm in cosmology since the beginning of
the 21st century.
Really, since the Dark energy (or $ \Lambda $-term in the General Relativity
equations) is distributed perfectly uniform everywhere, one should expect
that the cosmological expansion---at least, with a somewhat reduced
magnitude---takes place at any scale.

The degree of the corresponding reduction can be estimated
as~\cite{Dumin_08,Dumin_18}:
\beq
\label{H_loc-H}
\frac{H_0^{\rm (loc)}}{H_0} =
{\left[ 1 +
  \frac{\Omega_{{\rm D}0}}{\Omega_{{\Lambda}0}} \, \right]}^{-1/2} \! ,
\eeq
where $ \Omega_{{\Lambda}0} \! \approx \! 0.75 $ and
$ \Omega_{{\rm D}0} \! \approx \! 0.25 $~are the relative energy densities of
the Dark energy (or $ \Lambda $-term) and the irregularly-distributed
substance (mostly, the Dark matter)%
\footnote{
Strictly speaking, the ``global'' (intergalactic) Hubble constant~$ H_0 $
remains somewhat uncertain till now, which is often called the ``Hubble
tension''~\cite{Ryden_17}:
The various types of measurements resulted in the values approximately
from~67 to~73\,km/s/Mpc.
However, this discrepancy is below 10\,\% and, therefore,
rather insignificant as compared to the geophysical uncertainties plotted
in Fig.~\ref{fig:T-t}.
So, we can safely use everywhere some average value, \textit{e.g.},
$ H_0 \approx 70 $\,km/s/Mpc.}.
Therefore, we get:
$ {H_0} / {H_0^{\rm (loc)}} \! \approx 1.15 $, \textit{i.e.}, the Hubble
expansion should be suppressed locally by 15\,\%.
In fact, this is the lower limit.
In principle, $ H_0^{\rm (loc)} $ could be equal or even greater than~$ H_0 $
due to the contribution of the local matter---both dark and visible---to
the cosmological background density responsible for the formation of
the local expansion at the specified scale.
However, since the problem of averaging the cosmological background remains
a poorly understood subject till now, we shall refrain from any particular
estimates.

Let us mention that the value of~$ H_0^{\rm (loc)} $ reduced by~15\,\% as
compared to~$ H_0 $ is rather suitable to explain the anomalous recession
of the Moon from the Earth; \textit{e.g.}, our original
estimates~\cite{Dumin_02,Dumin_03}, which were subsequently refined
in~\cite{Dumin_08}.
Unfortunately, the even smaller value
$ H_0^{\rm (loc)} \! \approx 0.5\,H_0 \approx 0.35\,h $
(\textit{i.e.}, reduced by~50\,\%), which was originally proposed by Krizek
and Somer to keep the Earth's surface temperature permanent, turns out to be
inconsistent not only with the contemporary observational data but also with
the above-mentioned theoretical limit.
On the other hand, the substantially enhanced value
$ H_0^{\rm (loc)} \! \approx 2H_0 \approx 1.3{-}1.7\,h $,
which is required to fit the temporal evolution of the Earth's surface
temperature, does not contradict also the theoretical arguments.
Of course, a lot of work still needs to be done to specify this value more
accurately.
This should involve not only refinement of the Earth's surface temperature
in the past but also taking into account a number of additional physical
phenomena, such as the greenhouse effect, a variable albedo of the Earth,
\textit{etc.}

In summary, we believe that the Krizek--Somer anthropic principle, which is
based on resolution of the faint young Sun problem due to the local Hubble
expansion, is a very interesting theoretical concept.
However, the exact value of the local Hubble parameter appearing in this
mechanism remains rather uncertain.

\Acknow{YVD is grateful to
Yu.V.~Baryshev,
M.L.~Fil'chen\-kov,
S.M.~Kopeikin,
M.~K{\v{r}}{\'{\i}}{\v{z}}ek,
M.~Nowakowski, and
A.V.~Toporensky
for the discussions of the problem of local Hubble expansion.
Both authors are grateful to
E.G.~Khramova for the numerous consultations about the temperature on
the early Earth
as well as to O.~Primina for the support and encouragement of this work.}
The study was conducted under the state assignment of Lomonosov Moscow
State University.

\ConflictThey



\end{document}